\documentclass[aps,prl,twocolumn,groupedaddress,showpacs]{revtex4}
\usepackage{graphicx}

\begin{document}


\title{Speckle visibility spectroscopy and variable granular fluidization}


\author{P.K. Dixon$^{1,2}$ and D.J. Durian$^{1}$}
\affiliation{
    $^{1}$Department of Physics \& Astronomy, University of California,
    Los Angeles, CA 90095\break
    $^{2}$Department of Physics, California State University,
    San Bernardino, CA 92407
}


\date{\today}

\begin{abstract}
	We introduce a dynamic light scattering technique capable of
	resolving motion that changes systematically, and rapidly,
	with time.  It is based on the visibility of a speckle pattern
	for a given exposure duration.  Applying this to a vibrated
	layer of glass beads, we measure the granular temperature and
	its variation with phase in the oscillation cycle.  We observe
	several transitions involving jammed states, where the grains
	are at rest during some portion of the cycle.  We also observe
	a two-step decay of the temperature on approach to jamming.
\end{abstract}


\maketitle



A wealth of spectacular phenomena occur when granular materials are
subjected to periodic vertical vibration~\cite{jnb,duran}.  For
example, shallow layers exhibit period doubling and pattern
formation; deep layers exhibit heaping, convection,
and logarithmically-slow compaction.  It
is relevant to ask: What grain-scale physics gives rise to all this
intriguing macroscopic behavior?  Since the driving is generally at
high-amplitude but low-frequency, the dynamics vary dramatically
during the oscillation cycle.  At first, the grains are
jammed~\cite{cates,andreasid}, completely at rest in some random
packing configuration.  When the downward acceleration exceeds $-g$,
the grains are launched upward from the plate.  The layer
expands as the grains collide and move about randomly.  Soon they
crash back into the bottom plate, and pick up energy from this impact. 
Finally, they come to rest after rattling away energy by inelastic
collisions.  This approach to jamming can be accompanied by clustering
and ``inelastic collapse'', a finite time singularity where the
collision rate diverges and the collision length
vanishes~\cite{goldhirsch,du}.  This regime is especially significant
because granular hydrodynamics~\cite{jimstu,dufty,penn} and statistical
mechanics, extended to an athermal system, both break down.

Unfortunately, the above sequence of dynamics has not been
experimentally accessible.  Because of multiple light scattering,
video imaging is restricted to dilute granular gasses or surface
behavior~\cite{rouyer}.  Even then, the spatial resolution is much
larger than the collision length as the grains come to rest.  This
limitation holds for other imaging techniques as well, like
MRI~\cite{erichs,candela}, x-ray microtomography~\cite{seidler}, and
positron emission particle tracking~\cite{wildman02}.  Furthermore,
none is fast enough to capture the high collision rates when the
grains are barely fluidized.  By contrast, diffusing-wave spectroscopy
(DWS)~\cite{dws} is a dynamic light scattering (DLS)~\cite{berne}
method that applies to bulk granular media.  It has superior spatial
and temporal resolution, and can be extended to unsteady
dynamics~\cite{lemieux}.  However, it is based on temporal correlation
functions, which implicitly assume that all times are statistically
equivalent; therefore, it is not appropriate for periodic or aging
systems.  Altogether, the leading probe of dynamics in vibrated 3D
granular systems currently is an NMR technique~\cite{candela}.  It
allows tracking of individual grains throughout a highly-fluidized
sample to within about 150~$\mu$m and 1.4~ms.

In this paper, we introduce a new DLS technique and use it to study
the bulk grain dynamics throughout the oscillation cycle.  Here the
resolution is limited by the wavelength of light and by the speed of a
fast CCD camera.  Taking advantage of multiple light scattering, we
achieve a resolution of $\approx$1~nm and $\approx$20~$\mu$s.  With
this advance, we have unlocked regimes where the grains barely move. 
In particular, we observe three dynamic transitions: the onset of
fluidization, where the acceleration amplitude just exceeds $g$; a
jamming transition, where the grains crash into the plate and then
come to rest; and a transition to continuous fluidization, where the
grains do not jam up at any point during the cycle.  In contrast, the
NMR study of Ref.~\cite{candela} was conducted far above this point,
where granular hydrodynamics is applicable in the bulk.  Our three
transitions all involve a jammed state, and therefore cannot be
captured by granular hydrodynamics.  Since our observations quantify
the microscopics that underlie a host of intriguing phenomena, they
present a theoretical challenge.


Our granular system consists of $780\pm35~\mu$m diameter glass beads,
approximately twelve layers deep in a $10\times10$~cm$^{2}$ box with a
flat transparent bottom, vertical walls, and open top.  This is
mounted on a shake table, which in turn is leveled on an optical
bench.  Two three-axis accelerometers monitor the quality and peak
amplitude, $a_{\circ} \equiv \Gamma g$, of the vertical sinusoidal
oscillations.  All data are taken at frequency $f=10$~Hz.  We define
the phase to be $\phi=0$ when the plate is at height $z=0$ and moving
upward.  Properly leveled, we observe no discernable heaping, pattern
formation, or convection across the range of amplitudes studied,
$0<\Gamma<2.2$.

\begin{figure}
\includegraphics[width=3.00in]{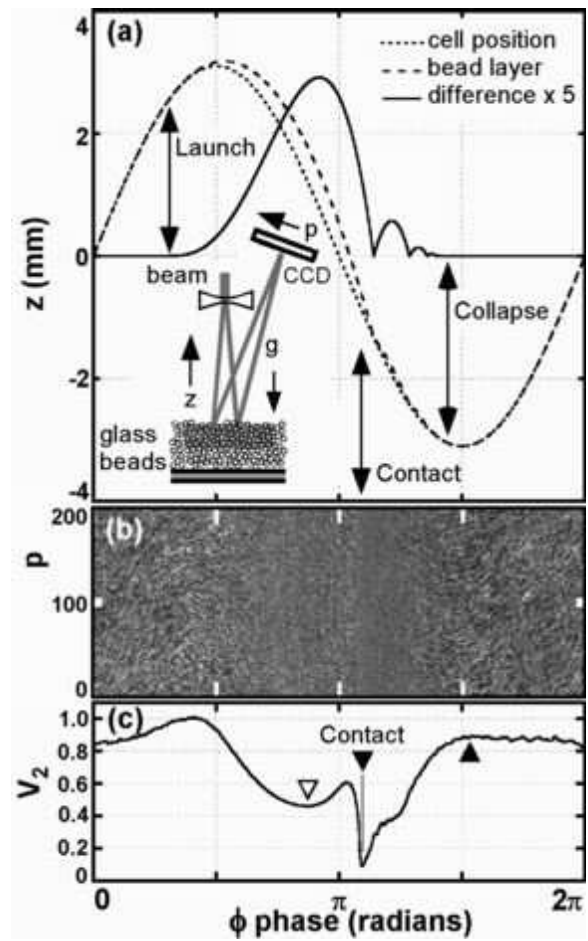}
\caption{(a) Theoretical position of the cell, bead layer, and their
difference ($5\times$), vs phase $\phi$ for a sinusoidally oscillated
cell.  The particular curves are for $\Gamma=1.25$ and $f=10$~Hz.  For
intuition, the bead layer is modeled as a slab with an effective
coefficient of restitution; a value of $\epsilon=0.5$ is used for the
data shown.  The inset shows the optical geometry.  (b) A
representative speckle image vs $\phi$ gathered by the linear CCD for
$\Gamma = 1.25$, $f = 10$~Hz, and $T=100~\mu$s.  For clarity, only 200
of 1024 pixels are shown.  (c) The averaged normalized variance
$V_{2}$ of 300 full images versus $\phi$ for the above settings.}
\label{setup}
\end{figure}

To measure grain motion, we introduce a method that we call ``speckle
visibility spectroscopy'' (SVS).  As shown in Fig.~\ref{setup}a, we
illuminate the grains from above with a $\approx$1~cm diameter beam of
a 100~mW frequency-doubled Nd-YAG laser ($\lambda=532$~nm). 
Photons perform a random walk with transport mean free path of
$\approx$4 grains.  Since the sample is 12 layers deep, about
${1\over3}$ of the photons reach the bottom plate, and the rest are
backscattered after a few scattering events.  The backscattered light
forms a speckle pattern in the far field, which we detect using a
digital linescan CCD camera (Basler-160: 1024 pixels, 8 bits deep),
and no intervening optics except a 532~nm filter.  The sample-CCD
distance is $\approx$20~cm, such that the speckle size is comparable
to the pixel size ($10~\mu$m).  Therefore, as the grains move and the
speckle pattern changes, large intensity fluctuations occur at each
pixel.  For the speckle pattern to be visible, however, the exposure
time of the CCD must be short compared to the time scale for speckle
fluctuations.  If the exposure time is longer, then the speckle blurs
out and the same average intensity is recorded at each pixel.  This is
the essence of SVS. It is illustrated qualitatively in
Fig.~\ref{setup}b by the space-time plot of speckle vs phase during
the cycle.  When the grains are at rest on the plate, the speckle is
clearly visible.  When the grains are fluidized, the speckle is
blurry.  The speckle is least visible just after impact, where rapid
grain motion is excited by the sudden injection of energy.  This is
reminiscent of ``laser speckle photography'', where the absence of
speckle in a laser-illuminated scene indicates
motion~\cite{briers,asakura}.

The key measurable quantity in SVS is the variance of intensity across
the pixels.  For an exposure duration $T$, each of the $N$ pixels
reports a time-integrated intensity, $S_{i}=\int_{0}^{T}I_{i}(t)dt/T$. 
The ensemble-averaged intensity and intensity-squared are computed as
$\langle I \rangle_{T}=\sum_{i=1}^{N}S_{i}/N$ and $\langle I^{2}
\rangle_{T}=\sum_{i=1}^{N}{S_{i}}^{2}/N$, respectively.  If there are
enough speckles, then the former is independent of $T$ and the
subscript may be dropped.  By contrast, $\langle I^{2} \rangle_{T}$
depends on $T$ and indicates the visibility of the speckles to the
extent that it exceeds $\langle I \rangle^{2}$.  To quantify
visibility on a scale of $0-1$, we define a normalized variance:
\begin{equation}
    V_{2}(T) \equiv \left[ \langle I^{2} \rangle_{T} /
    \langle I \rangle^{2} - 1 \right]/\beta.
\label{V2def}
\end{equation}
The factor $1/\beta$ is roughly the number of speckles per pixel, and
is determined experimentally by measuring the system at rest. 
Fig.~\ref{setup}c shows an example of $V_{2}(T)$ vs phase during the
cycle, for a fixed exposure of $T=100~\mu$s.  It is closer to 0 when
the grains are moving rapidly, and closer to 1 when the grains are
coming to rest.

To relate the variance to grain motion, note that
Eq.~(\ref{V2def}) involves {\it ensemble} averages, rather than time
averages.  Therefore, the Siegert relation~\cite{berne} holds, giving
$\langle I^{2}\rangle_{T} \equiv \langle
\int_{0}^{T}\int_{0}^{T}I_{i}(t')I_{i}(t'')dt'dt''/T^{2} \rangle_{i} =
\langle I \rangle^{2} \int_{0}^{T}\int_{0}^{T}[1+\beta
|g_{1}(t'-t'')|^{2}dt'dt''/T^{2}$.  Here, $\beta$ is the same as
above, and $g_{1}(t)$ is the normalized electric field
autocorrelation.  Since $g_{1}(t)$ is even, the double integral
simplifies and we arrive at the fundamental equation of SVS:
\begin{equation}
    V_{2}(T) = 2\int_{0}^{T}(1-t/T)|g_{1}(t)|^{2}dt/T.
\label{V2}
\end{equation}
The variance is thus a weighted average of $|g_{1}(t)|^{2}$ over the
exposure time $T$.  Both functions are 1 at short times, and 0 at long
times.  Given $g_{1}(t)$ from SVS measurements, the scattering site
motion may then be deduced by standard DLS practice.  For random
ballistic motion of average speed $\delta v$, the theory of DWS for
backscattered light gives $g_{1}(t)=\exp(-\gamma t)$ with $\gamma=4\pi
\delta v/\lambda$~\cite{dws}.  The corresponding variance is
\begin{equation}
    V_{2}(T)=2[\exp(-2\gamma T)-(1-2\gamma T)]/(2\gamma T)^{2}.
\label{gamma}
\end{equation}
At short times, the initial decay is linear: $V_{2}(T)\approx
1-{2\over3}\gamma T$; at long times, it is a power-law:
$V_{2}(T)\approx1/(\gamma T)$.  The heavy weighting in Eq.~(\ref{V2}) near
$t=0$ slows the decay, aiding in the measurement of fast
processes.

\begin{figure}
\includegraphics[width=3.00in]{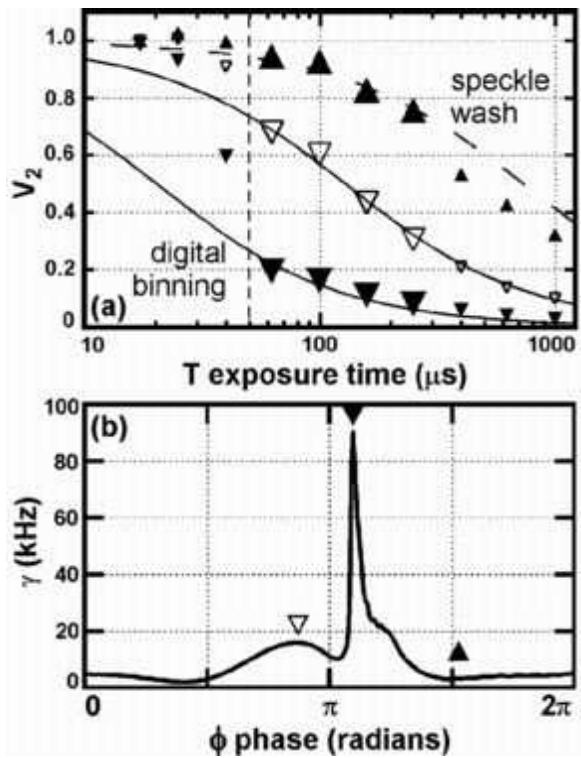}
\caption{(a) Semilog plot of $V_{2}$ vs exposure time $T$ for
$\Gamma=1.25$ and $f=10$~Hz at three phases $\phi$.  The
symbols correspond to the markers in Fig.~1c.  The curves are single
parameter fits using Eq.~(\ref{gamma}); only the data denoted by the
larger symbols has been fit.  The data fit by the dashed curve
represent the effect of speckle wash alone.  (b) The fitted rates vs 
phase.}
\label{V2T}
\end{figure}

We now return to Fig.~\ref{setup}, where the grains are vibrated at
$f=10$~Hz and $\Gamma=1.25$, and exploit our SVS method.  Three
representative phases in the cycle are marked by triangles in the
$V_{2}(100~\mu{\rm s})$ data in Fig.~\ref{setup}c: where the grain
motion is most rapid both in mid-flight and after impact, and where
the grain motion ceases.  For each of these three events, we show
$V_{2}(T)$ vs $T$ in Fig.~\ref{V2T}a.  All data tend toward 1 (0) at
short (long) times, as expected, where the speckle is most (least)
visible.  The actual dynamic range of our data is limited by two
effects that are specific to our experiment (not the SVS method). 
First, for exposures faster than $\approx 50~\mu$s, our 100~mW laser
produces a signal that is a small fraction of the 0-255 range of the
CCD. Therefore, the distribution of intensities is binned coarsely,
which systematically distorts the variance.  This limitation could be
reduced by a brighter laser.  Second, the
macroscopic motion of the system contributes to the variance, becoming 
significant for $T\rightarrow\infty$ and/or $V_{2}\rightarrow1$.  In
particular, even if the grains are at rest, the speckles form a static
pattern that washes as a whole across the CCD pixels.  This
``speckle-wash'' can be seen as a swirling in the space-time plot of
Fig.~\ref{setup}b.

The reliable portion of the variance data in Fig.~\ref{V2T}a compares
well with Eq.~(\ref{gamma}), as shown by the curves. 
The form of the decay is therefore
consistent with expectation: the grains appear to undergo random
ballistic motion with some average fluctuation speed $\delta v$.  For
granular materials, the kinetic energy associated with $\delta v$ is
called the granular temperature; it is directly proportional to the
decay rate $\gamma = 4\pi\delta v/\lambda$.  Results for $\gamma$ are
shown in Fig.~\ref{V2T}b as a function of phase in the cycle.  The
grains ``heat up'' both in mid-flight and from impact.  In both cases,
they lose energy through inelastic collisions.  About ${1\over5}$
cycle ($20$~ms) after impact, the grains lose their energy and
come to rest in a jammed state.  Due to speckle wash, however,
$\gamma$ doesn't decay to zero but rather to a readily-identifiable
baseline.  To our knowledge, this is the first measurement of the bulk
granular temperature during the collapse.

\begin{figure}
\includegraphics[width=3.00in]{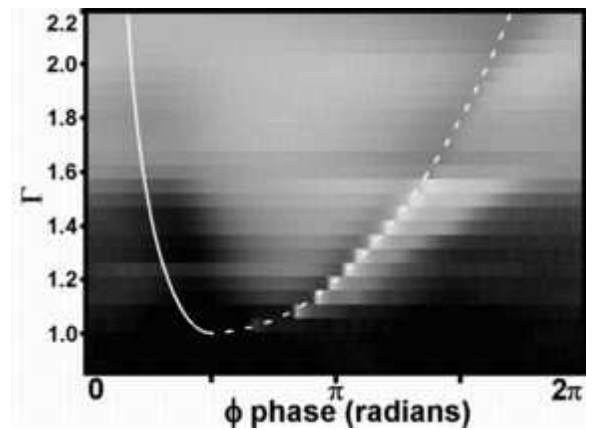}
\caption{Grayscale plot of decay rate $\gamma$, based on fits to
Eq.~(\protect\ref{V2}), as a function of both peak acceleration
$\Gamma$ and phase $\phi$ in the oscillation cycle; black=0,
white=120~kHz.  The solid white curve indicates where the grains are
launched, at an instantaneous acceleration of $-g$.  The dotted white
curve indicates when a free-fall object would land.}
\label{phase}
\end{figure}

We now repeat the experiment versus $\Gamma$, at constant $f=10$~Hz. 
A grayscale plot of $\gamma$ is shown in Fig.~\ref{phase} vs both
$\Gamma$ and $\phi$.  Effectively, this is a phase diagram denoting
the relative fluidization of the medium at different forcing rates and
at different points in the cycle.  Several transitions can be
observed.  First, below $\Gamma=1$, there is no fluidization at all. 
Above $\Gamma=1$, the grains become fluidized for some portion of the
cycle.  In particular, the grains are launched from the plate when the
instantaneous acceleration is $-g$, as denoted
by a solid-white curve.  The dotted-white curve denotes
where the grains crash back down, assuming they undergo simple
free-fall.  Some time after the grains land, they lose their energy
via collisions and jam.  The onset of both jamming and unjamming
depend on the driving amplitude.  For $\Gamma>1.6$, they merge
together and the system undergoes a transition to continuous
fluidization.  Though the grains never come to rest, their dynamics
still vary throughout the cycle.  As expected, the motion is fastest
(slowest) just after (before) impact.

\begin{figure}
\includegraphics[width=3.00in]{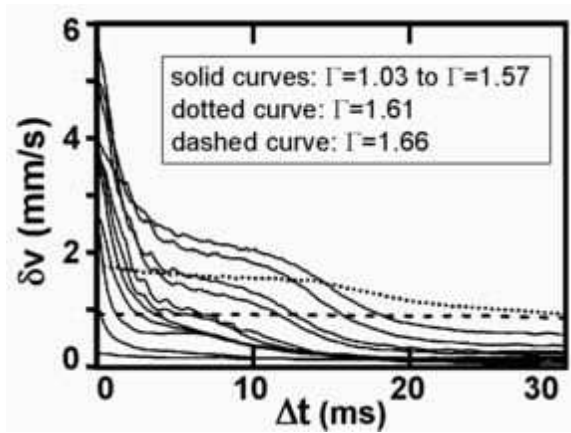}
\caption{Average fluctuation speed vs time after impact, extracted
from Fig.~\protect\ref{phase}, for different accelerations $\Gamma$.
For the solid curves, all below the transition to continuous
fluidization, $\delta v$ increases monotonically with driving
amplitude.  Above the transition, $\delta v$ never goes to zero and
doesn't spike as strongly when the grains impact the plate.}
\label{temperature}
\end{figure}

Both the process of jamming, and the nature of the transition to
continuous fluidization, may be studied in terms of the fluctuation
speed, $\delta v$, vs the time following impact, $\Delta t$.  These
data are extracted from Fig.~\ref{phase} and displayed in
Fig.~\ref{temperature}.  For amplitudes $1<\Gamma<1.6$, $\delta v$
spikes at impact and then decays to a plateau after $\approx$2~ms; the
initial decay rate appears to be independent of $\Gamma$.  Following
the plateau, $\delta v$ decays to zero (within resolution set by
speckle wash) as the grains jam up.  This second decay is slower than the
first.  The temperature spike, the level of the plateau, and the
duration of the plateau, all increase monotonically with $\Gamma$. 
The impacts become progressively more violent, until suddenly at
$\Gamma>1.6$ the plateau extends across the entire cycle and the
grains never come to rest.  The impacts are no longer as violent;
instead, the grains stay fluidized and $\delta v$ exhibits a smoother
variation with time.

\textit{Conclusion} The use of area detectors for ``multispeckle''
dynamic light scattering is on the rise~\cite{wong,luca,mochrie,pine}. 
The general approach has been to autocorrelate each pixel, and then to
average the results afterwards, all in software.  Our new method,
speckle variance spectroscopy (SVS), is dramatically different and
offers advantages in terms of both simplicity and
applicability.  In effect we have created a ``speckle ensemble
correlator'' in which all averages are explicitly-computed {\it
ensemble} averages of a single exposure of the CCD camera; neither
time averages, temporal autocorrelations, nor image storage are
necessary.  While the time resolution of prior methods is currently no
better than $\approx 2$~ms, we achieve $\approx 20~\mu$s.  More
significantly, we can follow dynamics that {\it change} on equally
rapid time scales.  This unprecedented resolution allows us to
capture the transition to continuous fluidization in Fig.~\ref{phase}
and the two-step decay of the granular temperature in
Fig.~\ref{temperature}.  With a brighter laser, and better control of
speckle wash, we are poised to study these and other granular dynamics
in greater detail.  SVS also opens a new window for the study of
bubble rearrangements in coarsening foams, motion and aging in glassy
suspensions, gelation, phase-separation, and other phenomena that
exhibit fast or quickly evolving nonstationary dynamics.

We gratefully acknowledge discussions with R. Ojha, P. Lemieux, and 
T. Usher, and the support of NSF-0070329.

\bibliography{svsrefs}

\end{document}